\documentclass{fic-l}

\theoremstyle{definition}

\theoremstyle{remark}

\numberwithin{equation}{section}

\newcommand{\Dslash}{D \! \! \! \! /} 
\newcommand{\kslash}{k \! \! \! /}

\newcommand{\partialslash}{\partial \! \! \! /} 
\newcommand{\half}{\mbox{\small{$\frac{1}{2}$}}} 
\newcommand{\Nc}{N_{\!c}} 
\newcommand{\Nf}{N_{\!f}} 
 
\newcommand{\NA}{N_{\!A}} 
\newcommand{\Nda}{N^d_{\!A}} 
\newcommand{\Noda}{N^o_{\!A}} 
\newcommand{\MSbar}{\overline{\mbox{MS}}} 

\usepackage{graphicx} 

\begin{document}

\title{Practicalities of Renormalizing Quantum Field Theories
} 

\author{J.A. Gracey} 
\address{Theoretical Physics Division \\ Department of Mathematical Sciences \\
University of Liverpool \\ P.O. Box 147 \\ Liverpool \\ L69 3BX \\ United
Kingdom} 



\begin{abstract}
We review the techniques used to renormalize quantum field theories at several 
loop orders. This includes the techniques to systematically extract the
infinities in a Feynman integral and the implementation of the algorithm 
within computer algebra. To illustrate the method we discuss the 
renormalization of $\phi^4$ theory and QCD including the application of the
critical point large $N$ technique as a check on the anomalous dimensions.
The renormalization of non-local operators in QCD is also discussed including
the derivation of the two loop correction to the Gribov mass gap equation in
the Landau gauge. 
\end{abstract}

\maketitle

\section{Introduction}

Our theoretical understanding of quantum phenomena in particle physics and
condensed matter is guided by the underlying quantum field theory. This is
formulated in terms of a Lagrangian of fields which have symmetries 
motivated by experimental observation. For instance, the electromagnetic
field governing light leads us to the gauge principle and gauge field theories, 
and in particular quantum electrodynamics. The properties of this field theory 
have been generalized to the current theory describing all elementary particles 
which is the standard model. In order to determine predictions from these 
quantum field theories requires one to develop the loop expansion of
perturbation theory where the interactions of the fields are represented by
Feynman diagrams. One of the main properties of such diagrams is that when
one evaluates them they are divergent. However, as has been established for a 
long time there is a systematic and mathematical procedure for handling the
resulting infinities which goes under the general title of renormalization. For
example, a comprehensive survey of the area is provided in \cite{1}. The topic 
is by nature a technical one and if one wishes to extract meaningful and 
accurate predictions from a quantum field theory, one needs to be able to 
perform the renormalization at a large number of loops. In this article we 
review some of the practicalities of achieving this in several field theories 
which are of main interest. These are scalar $\phi^4$ theory which is relevant 
to phase transitions in various spacetime dimensions and quantum chromodynamics
(QCD) which is the quantum field theory describing the quarks and gluons
associated with the strong nuclear force. Our aim is to address the basics in 
the first part of the article where we discuss various techniques for 
evaluating the complicated and divergent Feynman integrals which arise in 
renormalizable field theories, and the way they are handled in practice by 
using computer algebra and symbolic manipulation packages. Since the 
renormalization is a complicated procedure we also indicate the rudiments of 
the important task of how one actually verifies that a calculation is in fact 
correct. The second part of the article discusses recent problems of interest 
which apply the techniques discussed in the first part and illustrate their 
application from a practical point of view. These will primarily be in the 
context of QCD where, for instance, its three loop renormalization is discussed
in a particular non-linear gauge. We will also consider the problem of 
renormalizing operators which have a particular degree of non-locality.  

The article is organized as follows. We outline the main issues concerning
renormalization in section $2$ before discussing various calculational
techniques in section $3$. Section $4$ surveys the main checks one has on the
derivation of the renormalization group functions which lead on to more
specific checks for QCD in the large $\Nf$ limit in section $5$. We discuss
more recent aspects of renormalization of QCD in section $6$, before examining
the issue of how one can treat a particular class of non-local operators in
section $7$. Finally, various concluding remarks are provided in section $8$. 

\section{Statement of problem} 

We commence by summarizing from a general point of view the main issues 
underlying the renormalization of a quantum field theory and the main
terminology of the subject used in this article. At the outset it is worth
doing this with several basic field theories in mind but initially we will
focus on the scalar quartic interaction, $\phi^4$, in four spacetime 
dimensions. Later we will consider theories with gauge symmetries and in 
particular QCD. However, the general remarks and comments will apply equally to
all field theories which are renormalizable in four or other dimensions. For 
$\phi^4$ theory the basic Lagrangian is 
\begin{equation}
L ~=~ \frac{1}{2} \left( \partial \phi_o \right)^2 ~+~ \frac{g_o}{4!} \phi_o^4 
\label{phi4lag} 
\end{equation}
where the subscript ${}_o$ denotes that the field $\phi$ and coupling constant 
$g$ are {\em bare} quantities. This observation is essentially the foundation 
of the problem of the need for renormalization. If one were to naively compute 
quantum corrections to any Green's function with (\ref{phi4lag}) as the 
starting point, then the resulting Feynman diagrams would be infinite in four 
dimensions. This is not surprising as one is using a (local) Lagrangian 
involving quantum fields defined at the same spacetime point. In other words 
the variables, parameters or fields of the initial Lagrangian can be regarded 
as being insufficient or not the correct variables in which to define the 
problem. To circumvent this, one defines a new set of (renormalized) variables 
by scaling the original variables with a (multiplicative) factor, known as the 
{\em renormalization constant}. For (\ref{phi4lag}) we define these formally as
$\phi_o$~$=$~$\sqrt{Z_\phi} \phi$ and $g_o$~$=$~$Z_g g$ which leads to the 
Lagrangian in terms of renormalized variables 
\begin{equation} 
L ~=~ \frac{1}{2} Z_\phi \left( \partial \phi \right)^2 ~+~ \frac{g}{4!}
Z_\phi^2 Z_g \phi^4 ~. 
\end{equation}
This establishes the framework for renormalization. If one subsequently 
computes the previously infinite Green's functions with these, as yet 
undetermined renormalization constants, then by choosing their value 
appropriately the infinities can be absorbed into the renormalization 
constants, $Z_\phi$ and $Z_g$. As the quantum field theory is renormalizable in
four spacetime dimensions then the choice one makes for say the divergent 
$2$-point and $4$-point Green's functions of $\phi^4$ theory, means that
{\em all} other $n$-point functions are finite. Non-renormalizable theories 
require additional operators in the Lagrangian over and above the original ones
to retain finiteness. By contrast a superrenormalizable theory either does not
require all the available renormalization constants or these actually have a
finite number of terms when expanded in a coupling constant expansion. Whilst 
this is the general procedure, one ordinarily fixes the renormalization 
constants order by order in a coupling constant or perturbative expansion. 
Though one can still follow this prescription non-perturbatively such as in a 
lattice regularization. 

Whilst this is the overall essence of renormalization, there are clearly
several technical issues to be addressed which we now briefly discuss. The
choice in redefining the variables may seem ad hoc and has little connection
with the real physical world. Indeed one may not only absorb the infinities of 
the Green's functions but also an arbitrary finite part. How one practically 
removes the divergences is known as the renormalization scheme. Although there 
are a large number of such schemes the most widely used is the modified minimal
subtraction scheme denoted by $\MSbar$. To ensure that the same physics emerges
independently of how the infinities are absorbed, the information contained in 
the renormalization constants are encoded in renormalization group functions 
such as $\gamma_\phi(g)$ and $\beta(g)$ which appear in the renormalization 
group equation. These functions determine properties of the quantum theory. For
instance, in field theories underlying condensed matter problems the critical 
points relating to the phase transitions are given by the non-trivial values of
the coupling constant, $g_c$, where $\beta(g_c)$~$=$~$0$. In particle physics 
applications, for example, the solution of the differential equation defined by
the $\beta$-function determines how the running coupling constant depends on 
the renormalization scale, which is defined later. For instance, in QCD since 
the $\beta$-function is negative for small values of the coupling constant then
the theory is asymptotically free, \cite{2}. In other words the (confined) 
quarks and gluons behave at very large energies as if they were free particles.

In alluding to infinities in general terms so far, for practical calculations
one must have a mathematical way of handling them and determining their nature.
Moreover, for theories with a classical symmetry which is assumed to be 
preserved in the quantum theory, the procedure for quantifying or {\em 
regularizing} the infinities must respect the symmetries of the theory. In this
context we briefly mention the algebraic renormalization technology developed 
primarily to determine the general consequences for the renormalization of 
gauge theories but also for theories with supersymmetry. For example, see 
\cite{3}. In essence algebraic renormalization determines the form of the 
renormalization constants consistent with the Slavnov-Taylor identities.
Two of the most widely used regularization procedures are lattice
regularization and dimensional regularization. In the former one replaces or
approximates the continuum of spacetime by a discrete set of points regularly
spaced a distance $a$ apart\footnote{This is known as a regular or square
lattice which is the most widely used. However, lattice regularization is not
restricted to square lattices. One can define triangular or random lattices.}.
The presence of $a$ is a feature of any regularization which is that every
regularization introduces an arbitrary scale. For lattice regularization one
wishes to take $a$ as small as possible in order to be as close as possible to 
the continuum. Though in practical (and financial) terms this is very costly 
from the point of view of computer running time. However, the point is that 
whatever physical quantity one is computing the result has to be independent of
the arbitrary regularization scale. 

The same feature is true in the other regularization we consider. By contrast, 
one retains a continuum spacetime but changes the dimension away from four by 
an infinitesimal amount by defining the spacetime dimension as 
$d$~$=$~$4$~$-$~$2\epsilon$ where $\epsilon$~$\ll$~$1$ and $\epsilon$~$>$~$0$. 
In this regularization the infinities in Green's functions manifest themselves 
as poles in $\epsilon$ and after the poles are removed by some criterion in 
some scheme, then the $\epsilon$~$\rightarrow$~$0$ limit can be taken in a 
non-singular way. However, in changing the spacetime dimension an arbitrary 
scale is introduced by requiring that the coupling constant remains 
dimensionless in $d$-dimensions. We now take
\begin{equation}
g_o ~=~ \mu^{2\epsilon} Z_g g
\end{equation}
where $\mu$ is the arbitrary renormalization scale. Since the Green's function 
after renormalization, $\Gamma^{(n)}(\mu,p,g,\ldots)$, will necessarily depend 
on $\mu$ in the combination $p^2/\mu^2$ where $p$ is a momentum, we can 
quantify the essence of the renormalization group as 
\begin{equation}
\mu \frac{d~}{d\mu} \Gamma^{(n)}(\mu,p,g,\ldots) ~=~ 0 
\end{equation}
so that results are independent of the arbitrary scale $\mu$.

One point of clarification needs to be made in the context of defining
(dimensional) regularization which is that infinities fall into two classes.
Those which are ultraviolet arising from divergences in Feynman integrals at 
large momenta and those which are infrared coming from divergences in integrals
at low loop momenta. In massless theories dimensional regularization 
regularizes both types of infinities so that it is never clear where the poles 
in $\epsilon$ originate from. However, infrared infinities can be regularized 
by a non-zero mass which clearly acts as a low momentum cutoff. This mass can 
either be present in the original Lagrangian or introduced by hand. In the 
latter case when the theory is renormalized it can be smoothly set to zero. 
However, in gauge and supersymmetric theories such extra masses can break the 
symmetries and if such regularizing masses are introduced, care is required to 
preserve the symmetry.   

We close this section by briefly summarizing several renormalization schemes
which are in common use. We have alluded to $\MSbar$ already and we note that
the original minimal subtraction scheme, MS, on which it is based requires
the removal of the poles in $\epsilon$ only. (Though it is worth noting that
MS and $\MSbar$ are not solely tied to dimensional regularization. One can
minimally subtract divergences in lattice regularization.) The $\MSbar$ scheme
is a modification of MS in that a common finite term, $4\pi e^{-\bar{\gamma}}$
where $\bar{\gamma}$ is the Euler-Mascheroni constant, is removed in addition 
to the poles as it was observed that in practical calculations that the
convergence of perturbative expansions was improved, \cite{4}. These two 
schemes fall into the class of mass independent renormalization schemes. By 
contrast mass dependent schemes are those where, aside from the poles in 
$\epsilon$ being removed certain finite parts are also removed which involve 
masses of the fields introduced by putting the external particle masses 
on-shell or setting a Green's function to a particular mass independent number 
at some value of the renormalization scale. Examples of such mass dependent 
schemes are MOM, \cite{5}, on-shell and the modified regularization 
(RI$^\prime$), \cite{6}, schemes. In lattice regularization RI$^\prime$ is one
of the main choices as it is constructed in such a way as to minimize computer 
calculation time. In a similar way to requiring that physical quantities are 
independent of the renormalization scale introduced through the regularization,
they also have to be independent of the renormalization scheme. However, it is 
possible to convert between different renormalization schemes. Indeed the 
$\MSbar$ scheme is used as the reference in this context primarily because one 
can compute much further in the coupling constant expansion than in a mass 
dependent renormalization scheme. Therefore, on the assumption that with more 
terms the series is more accurate, it should be closer to the real physical 
value of the quantity being computed. So, for example, in lattice computations 
which are non-perturbative but computed in the RI$^\prime$ scheme, one has to 
convert the results to the $\MSbar$ scheme and match the large momentum 
behaviour to the perturbative $\MSbar$ result at whichever perturbative order 
is available. Thus the larger the number of terms in the coupling constant 
expansion that are available then this will reduce the error in the final 
numerical estimate. For certain problems this has been achieved at three and 
four loops, \cite{7,8}.  

\section{Calculational methods} 

Having discussed the method of renormalization in general terms, we now detail
some of the techniques used to extract regularized infinities from Feynman
integrals. We will concentrate on the application of dimensional regularization
to massless and massive Feynman diagrams. Throughout we will regard the 
integrals as expanded near four dimensions with $d$~$=$~$4$~$-$~$2\epsilon$. 
First, we recall a simple but powerful technique to treat one loop diagrams 
which is based on a simple integral and attributed to Feynman. Within a Feynman
graph one has products of propagators and the basic idea is to write these as 
an integral using 
\begin{equation}
\frac{1}{ab} ~=~ \int_0^1 \frac{dx}{[ax + (1-x)b]^2} 
\end{equation} 
where notionally $a$ $=$ $k^2$ $-$ $m_1^2$ and $b$ $=$ $(k-p)^2$ $-$ $m_2^2$ 
with $k$ regarded as an internal momentum and $p$ as external. With the two 
momenta within one factor, one can use Lorentz symmetry before performing the
$k$-integration. This leaves a function of the Feynman parameter, $x$, which at
one loop can be related to known functions. For example, when the masses $m_1$ 
and $m_2$ are equal the integral can be written in terms of a hypergeometric 
function. Whilst this is a powerful approach at one loop, it ceases to be 
practical at higher loops especially when there are more than one mass and 
external momenta. In practice one chooses a method of calculating the Feynman 
integral which is tailored to the overall (renormalization) problem of 
interest.

We will focus on one of these and consider the situation where the field theory
involves particles of one non-zero mass and is renormalizable. Then to 
renormalize the divergent Green's functions and to extract the infinities, one 
expands the set of (scalar) Feynman integrals in powers of $p^2$ where $p$ is 
the external momentum. This method is known as the vacuum bubble expansion. In 
regarding all integrals as scalars, we have assumed that one has first
decomposed Lorentz tensor integrals into their scalar pieces. To focus on the 
particulars of the vacuum bubble expansion we consider the simple example of a 
self energy type integral in a $2$-point function which gives  
\begin{eqnarray}
\int_k \frac{1}{[k^2-m^2][(k-p)^2-m^2]} &=& \int_k \frac{1}{[k^2-m^2]^2} ~-~ 
\int_k \frac{p^2}{[k^2-m^2]^3} \nonumber \\
&& ~+~ \frac{4}{d} \int_k \frac{k^2p^2}{[k^2-m^2]^4} ~+~ O((p^2)^2) ~. 
\label{vacbubexp}
\end{eqnarray}
In the final term the $(kp)^2$ numerator factor has been simplified by using
Lorentz symmetry. The expansion truncates due to the renormalizability 
condition which states that since the theory is renormalizable the $O((p^2)^2)$
terms are finite since otherwise one would require a fourth derivative
$2$-point term which is forbidden by renormalizability. The resulting integrals
on the right hand side of (\ref{vacbubexp}) are simple vacuum loops which can
be evaluated. More appropriately the method can be easily extended to higher
loops. For instance, in four dimensions three loop single mass scale vacuum 
bubbles are known to their finite part and at two loops, three mass scale 
vacuum graphs are also known to their finite part with respect to $\epsilon$. 
See, for example, \cite{9}. One of the benefits of this algorithm is that it 
can be implemented in computer algebra in an automatic way. Also the method is 
applicable to higher leg Green's functions if one wants to renormalize them
too. However, in certain problems where the finite part is required exactly
in, say, scattering problems, the more recent technique of \cite{10} is
appropriate. There each propagator is represented by a Mellin-Barnes integral
and contour integration used to evaluate the two loop $4$-point functions, for
example, to their finite parts. Though currently results have yet to be derived
for the physically relevant case of all possible massive propagator 
combinations.  

Next, we note that one distinct advantage of the massive vacuum bubble
expansion is that there are no infrared problems as the inherent mass provides
a {\em natural} infrared regulator. For theories where there are massless
fields it may seem that the vacuum bubble approach is inapplicable. However,
\cite{11}, one can manufacture a fictitious mass $\bar{\mu}$ which acts as an 
infrared regulator via the identity  
\begin{equation}
\frac{1}{(k-p)^2} ~=~ \frac{1}{[k^2-\bar{\mu}^2]} ~+~ 
\frac{[2kp - p^2 - \bar{\mu}^2]}{(k-p)^2[k^2-\bar{\mu}^2]} ~.
\end{equation}
Within a Feynman diagram this identity can be used repeatedly with the 
truncation criterion based on Weinberg's theorem, \cite{12}, for the overall
finiteness of a Feynman integral. Like the completely massive case this
algorithm can be implemented automatically in computer algebra. Though one
ought to be aware of the potential problem of breaking an inherent symmetry of 
the theory when a non-zero $\bar{\mu}$ is introduced. For non-abelian gauge 
theories this is discussed in \cite{11}.

Another equally powerful approach for massless field theories, such as QCD,
is the use of integration by parts based on the identity, in $d$-dimensions,
\cite{13}, 
\begin{equation}
0 ~=~ \int \frac{d^dk}{(2\pi)^d} \frac{\partial ~}{\partial k^\mu} \left[ k^\mu
I(p,k,....) \right] 
\end{equation}
where $I(p,k,....)$ is the integrand derived from the propagators and vertices.
Although such identities can be used for massive integrals, the power lies in 
the observation that the differentiation can introduce numerator propagator 
factors to simplify the momentum topology. By momentum topology we mean that
diagram which represents all the scalar propagators. This is not necessarily
the same as the actual Feynman diagram topology itself as the cancellation of a
denominator factor means that that line would be omitted in the momentum
topology. This cancellation is one of the principles which underlies the 
{\sc Mincer} algorithm, \cite{14}. This is a package for evaluating massless 
three loop $2$-point functions in dimensional regularization to their finite 
parts. Moreover, it has been encoded in the symbolic manipulation language 
{\sc Form}, \cite{15,16}. Essentially at three loops there are fourteen basic 
integration topologies each with its own integration by parts routine. Though 
these momentum topologies are not all independent. Whilst primarily used for 
$2$-point functions, {\sc Mincer} can be applied to $3$-point functions when 
one of the external momenta is set to zero or nullified. However, in this case 
one must ensure that the nullification does not introduce spurious infrared 
singularities such as $\int_k \frac{1}{(k^2)^2}$ which result in poles in 
$\epsilon$ which being infrared in nature, cannot be distinguished from 
ultraviolet ones. Despite this technicality there are methods for handliing 
nullification known as infrared rearrangement, \cite{17}. Though this has not 
been implemented automatically in computer algebra. The main application of 
{\sc Mincer} is to the renormalization of four dimensional gauge theories at 
three loops and in particular QCD.

Given this emphasis on automatic computer algebra we make some specific
remarks. The need for such machinery can be gauged from the fact that when one
increases the loop order of a calculation, the number of Feynman diagrams
increases almost exponentially. As examples we note that the recent full
three loop renormalization of QCD in the maximal abelian gauge required the
evaluation of $37322$ diagrams, \cite{18}. Also the four loop QCD 
$\beta$-function of \cite{19} in a linear covariant gauge required of the order
of $50000$ diagrams. Clearly computers are necessary to not only implement the 
computational algorithm, such as {\sc Mincer} or the vacuum bubble method, but 
also to handle the sum of the individual results. Packages such as 
{\sc Mincer}, which have been optimized, are essential to having as short a 
computation time as possible. Indeed a publicly available four loop 
{\sc Mincer} would be equally as useful. For such automatic computations the 
Feynman diagrams themselves need to be generated electonically and {\sc Qgraf},
\cite{20}, has been developed specifically for this purpose. It has various 
output formats which can be readily converted to the notation used by {\sc 
Mincer} before applying the algorithm itself. The implementation of the 
renormalization procedure can also be performed automatically without the need 
for the traditional method of subtractions. This method determines the absolute 
divergence of a diagram by subtracting all subgraph divergences. Instead in the
automatic approach, \cite{21}, the Green's function is computed as a function 
of the {\em bare} parameters. Then the counterterms are introduced naturally 
(and equivalently to the subtraction method) by rescaling by the perturbatively
expanded renormalization constants $g_o$~$=$~$Z_g g$ and so on. Once the 
counterterms have been implemented at a particular loop order, the divergence 
remaining is then that associated with the renormalization constant of that 
particular Green's function. 

Finally, we briefly comment on other recent techniques of evaluating Feynman 
diagrams. One which is also appropriate to $n$-point functions at two loops for
non-zero external momenta is the differential equation method of \cite{22}. The
basic idea is to derive a complete set of differential equations at a
particular loop order for relevant momentum topologies. These are then solved
with a basic set of master integrals as the boundary conditions which have to 
be evaluated by direct methods. For completely massless integrals in problems 
where conformal symmetry is present, such as at a fixed point, the method of 
uniqueness, \cite{23,24}, is also powerful. Essentially when the sum of powers 
of three momenta in a loop integral satisfy a particular condition depending on 
the spacetime dimension, then the integral can simply be replaced by products 
of related propagators and Euler $\Gamma$-functions of the original propagator 
exponents. For large $\Nf$ methods, which will be discussed later, this has 
proved to be extremely powerful in computing information on the 
renormalization group functions beyond the leading large $\Nf$ order and to
all orders in perturbation theory. 

\section{Checks} 

Performing a renormalization even of a simple quantum field theory to several 
loop orders can lead to the computation of a large number of Feynman diagrams 
which are determined by one or other of the methods previously discussed. 
However, one natural question arises in the ultimate derivation of the 
renormalization group functions and that is whether the results are correct. 
For major loop calculations in four dimensional gauge theories to check by 
repeating the work using independent computer algebra programmes will not only 
be time consuming but make a large demand in both human and computer resources,
which could be used for other problems. The exception to this is the situation 
where one is extending existing anomalous dimensions and $\beta$-functions to 
the next loop order. This requires the counterterms which depend on the finite 
part of the Green's functions which were renormalized in the previous loop 
calculation but which are not ordinarily computed to construct the established 
renormalization constants. Therefore, one invariably has to reconstruct the 
previous calculation prior to tackling the new loop order. Aside from this 
check we now address one standard way of assessing whether the calculation one 
has performed is correct. By this we mean that it satisfies a set of internal 
consistency checks at the very least and others tailored to the problem in 
hand. First, the renormalization group provides a clear insight into the 
structure of the renormalization constants. To illustrate this, if we work from
the form of the $\MSbar$ $\beta$-function and anomalous dimension 
$\gamma_\phi(g)$ given by 
\begin{eqnarray} 
\beta(g) &=& (d-4)g + A g^2 + B g^3 + C g^4 + O(g^5) \nonumber \\
\gamma_\phi(g) &=& a g + b g^2 + c g^3 + O(g^4) 
\end{eqnarray} 
in $d$ $=$ $4$ $-$ $2\epsilon$ dimensions where $g$ is a generic coupling
constant then the corresponding renormalization constants must be of the form
\begin{eqnarray} 
Z_g &=& 1 - \frac{Ag}{(d-4)} ~+~ \left( \frac{A^2}{(d-4)^2} - \frac{B}{2(d-4)}
\right) g^2 \nonumber \\
&& +~ \left[ - \frac{A^3}{(d-4)^3} + \frac{7AB}{6(d-4)^2} - \frac{C}{3(d-4)}
\right] g^3 + O(g^4) \nonumber \\  
Z_\phi &=& 1 + \frac{ag}{(d-4)} ~+~ \left( \frac{(a^2-aA)}{2(d-4)^2} 
+ \frac{b}{2(d-4)} \right) g^2 \nonumber \\
&& +~ \left[ \frac{(2aA^2-3a^2A+a^3)}{6(d-4)^3} 
+ \frac{(3ab-2aB-2bA)}{6(d-4)^2} - \frac{c}{3(d-4)} \right] g^3 \nonumber \\
&& +~ O(g^4) 
\end{eqnarray} 
to three loops. Clearly the residues of the simple poles are in a one to one
correspondence with the coefficients of the renormalization group functions
since the $\MSbar$ scheme is used. However, the residues of the other poles
depend only on the structure of the {\em previous} loop calculations. 
Therefore, in a new loop calculation these poles are already predetermined and 
{\em must} emerge from the new one loop calculation. This is important in 
automatic calculations since one determines the full renormalization constant
without using the subtraction method. Another internal check is that provided 
by symmetries of the original theory. Ignoring the technicalities produced by 
anomalies, in a gauge theory certain renormalization group functions are 
independent of the choice of gauge in mass independent renormalization schemes.
Therefore, working in arbitrary covariant gauges means that the gauge parameter
must be absent in the final result. By contrast, when renormalization group 
functions which depend on the gauge parameter, $\alpha$, are calculated, the 
residues of the triple and double poles of the renormalization constants will 
depend on $\alpha$ and also be constrained by the conditions above. Briefly, in 
supersymmetric theories similar conditions emerge. For example, when 
supersymmetry is unbroken, the anomalous dimensions of the fields in the same 
supermultiplet have to be equal when the component field version of the 
Lagrangian is considered. This is a stringent check not only on the 
renormalization procedure but also on the regularization which must preserve 
the supersymmetry. Whilst this ensures the renormalization constants are 
checked to an extent, the residues of the simple poles are not. However, there 
are partial checks available in some theories from the large $\Nf$ methods 
developed for scalar field theories, \cite{24,25,26}.

This can be illustrated in the case of $O(N)$ $\phi^4$ theory where the scalar
field is a vector in $O(N)$. In $\MSbar$ in $d$-dimensions the $\beta$-function
has the form 
\begin{eqnarray} 
\beta(g) &=& (d-4)\frac{g}{2} ~+~ [N+8] \frac{g^2}{6} ~-~ [3N+14] \frac{g^3}{6}
\nonumber \\ 
&& +~ \left[33N^2 + 922N + 2960 + 96(5N+22)\zeta(3) \right] 
\frac{g^4}{432} ~+~ O(g^5) 
\label{phi4beta}
\end{eqnarray}  
where $\zeta(n)$ is the Riemann zeta function. In $d$-dimensions there is a
non-trivial fixed point of the $\beta$-function, $g_c$, defined by 
$\beta(g_c)$~$=$~$0$ known as the Wilson-Fisher fixed point. Clearly $g_c$ will
be a function of $d$ and $N$. Though near four dimensions with 
$d$~$=$~$4$~$-$~$2\epsilon$ we have $g_c$~$=$~$g_c(\epsilon,N)$, which can be 
expanded in powers of $1/N$ as $N$~$\rightarrow$~$\infty$ giving  
\begin{equation}
g_c ~=~ \frac{6\epsilon}{N} ~+~ \left[ \, -~ 48 \epsilon + 108 \epsilon^2 
- 99 \epsilon^3 + O(\epsilon^4) \right] \frac{1}{N^2} ~+~ O\left( \frac{1}{N^3} 
\right) ~. 
\end{equation} 
Evaluating the renormalization group invariant universal critical exponent
$\beta^\prime(g_c)$ in the same limit gives 
\begin{equation}
\beta^\prime(g_c) ~=~ - \epsilon ~+~ \left[ 
18 \epsilon^2 - 33 \epsilon^3 - \frac{5}{2} \epsilon^4 + O(\epsilon^5) \right] 
\frac{1}{N} ~+~ O\left( \frac{1}{N^2} \right) 
\end{equation} 
which encodes the information of the original $\beta$-function, albeit in a
different way. However, if one can compute $\beta^\prime(g_c)$ in
$d$-dimensions order by order in the $1/N$ expansion then the coefficients in
the polynomials of $N$ in (\ref{phi4beta}) can be read off directly. In 
\cite{24,25,26,27} the $d$-dimensional large $N$ technique was developed for 
the $O(N)$ non-linear $\sigma$ model in $d$~$=$~$2$~$+$~$\varepsilon$ 
dimensions and the first three terms in the $1/N$ series for the exponents 
$\eta$, $\nu$ and $\omega$ have been determined in $d$-dimensions. As this 
model is in the same universality class as $O(N)$ $\phi^4$ theory at the 
Wilson-Fisher fixed point then one can partially check off the known 
coefficients in the corresponding renormalization group functions. For 
instance, writing the leading large $N$ wave function anomalous dimension as 
\begin{equation} 
\gamma(g) ~=~ \sum_{r=1}^\infty ( c_r N^2 + d_r N + e_r ) N^{r-2} g^{r+1} 
\end{equation} 
then with $U_{6,2}$~$=$~$\sum_{n>m>0}^\infty \frac{(-1)^{n-m}}{n^6 m^2}$ as 
the $(3,4)$ torus knot number, \cite{28}, one finds at {\em ten} loops and 
$O(1/N^3)$ that 
\begin{eqnarray} 
e_9 &=& [1560674304 \zeta(10) ~-~ 12534896640 \zeta(9) ~+~ 11070010560 
\zeta(8) \nonumber \\
&& +~ 1732018176 \zeta(7) \zeta(3) ~+~ 581961984 \zeta(7) ~-~ 
3411394560 \zeta(6) \zeta(3) \nonumber \\
&& -~ 2684240640 \zeta(6) ~+~ 209534976 \zeta^2(5) ~-~ 
1567752192 \zeta(5) \zeta(4) \nonumber \\
&& +~ 1754664960 \zeta(5) \zeta(3) ~-~ 975533568 \zeta(5) ~-~ 
9289728 \zeta(4) \zeta^2(3) \nonumber \\
&& +~ 1310201856 \zeta(4) \zeta(3) ~+~ 1636615872 \zeta(4) ~-~ 
137158656 \zeta^3(3) \nonumber \\
&& -~ 1708996608 \zeta^2(3) ~+~ 294403968 \zeta(3) \nonumber \\
&& -~ 89800704 U_{62} ~-~ 341350433]/1950396973056 ~. 
\end{eqnarray} 
This requires not only knowledge of $\eta$ at $O(1/N^3)$ but also $\omega$ at
$O(1/N^2)$ as this encodes the value of $g_c$ required for 
$\eta$~$=$~$\gamma(g_c)$. By way of illustration as to the form of such large
$N$ exponents we note that with 
\begin{equation}
\omega ~=~ \mu ~-~ 2 ~+~ \sum_{i=1}^\infty \frac{\omega_i}{N^i}  
\end{equation} 
then 
\begin{equation} 
\omega_1 ~=~ (2\mu-1)^2\eta_1    
\end{equation} 
and, \cite{27},  
\begin{eqnarray} 
\omega_2 &=& \left[ -~   \frac{4(\mu^2-5\mu+5)(2\mu-3)^2(\mu-1)\mu^2 ( 
\bar{\Phi}(\mu) + \bar{\Psi}^2(\mu) )} {(\mu-2)^3(\mu-3)} \right. \nonumber \\ 
&&-~ \left. \frac{16\mu(2\mu-3)^2}{(\mu-2)^3(\mu-3)^2\eta_1} 
\right. \nonumber \\  
&&-~ \left. \frac{3(4\mu^5-48\mu^4+241\mu^3-549\mu^2+566\mu-216)(\mu-1)\mu^2 
\hat{\Theta}(\mu)}{2(\mu-2)^3(\mu-3)} \right. \nonumber \\ 
&&-~ \left. [16\mu^{10}-240\mu^9+1608\mu^8-6316\mu^7+15861\mu^6 \right. 
\nonumber \\ 
&&~~~~ \left. -~ 25804\mu^5+26111\mu^4-14508\mu^3+2756\mu^2 \right. 
\nonumber \\
&&~~~~ \left. +~ 672\mu-144)]/[(\mu-2)^4(\mu-3)^2] \bar{\Psi}(\mu) \right. 
\nonumber \\ 
&&+~ \left. [144\mu^{14}-2816\mu^{13}+24792\mu^{12}-130032\mu^{11} 
+452961\mu^{10} \right. \nonumber \\ 
&&~~~~ \left. -~ 1105060\mu^9+1936168\mu^8-2447910\mu^7+2194071\mu^6 \right. 
\nonumber \\ 
&&~~~~ \left. -~ 1320318\mu^5+460364\mu^4-43444\mu^3-26280\mu^2 \right. 
\nonumber \\ 
&&~~~~ \left. +~ 8208\mu-864]/[2(2\mu-3)(\mu-1)(\mu-2)^5(\mu-3)^2\mu] \frac{}{}
\right] \eta_1^2  
\end{eqnarray} 
where $\eta_1$ $=$ $-$ $4\Gamma(2\mu-2)/[\Gamma(2-\mu)\Gamma(\mu-1)
\Gamma(\mu-2)\Gamma(\mu+1)]$ and we have set $d$~$=$~$\mu/2$. The expression 
for $\omega_2$ contains derivatives of the $\Gamma$-function which are denoted 
by
\begin{eqnarray} 
\bar{\Psi}(\mu) &=& \psi(2\mu-3) ~+~ \psi(3-\mu) ~-~ \psi(\mu-1) ~-~ \psi(1)
\nonumber \\
\hat{\Theta}(\mu) &=& \psi^\prime(\mu-1) ~-~ \psi^\prime(1) \nonumber \\
\bar{\Phi}(\mu) &=& \psi^\prime(2\mu-3) ~-~ \psi^\prime(3-\mu) ~-~ 
\psi^\prime(\mu-1) ~+~ \psi^\prime(1) ~.
\end{eqnarray}  
Whilst $O(N)$ $\phi^4$ theory has been examined to several orders in $1/N$,
the same technique has been developed for QCD in \cite{29,30}. There the 
expansion is in terms of the number of quark flavours, $\Nf$, and {\em not} 
$\Nc$ which is the number of colours. The latter $1/\Nc$ expansion deals with 
the structure of QCD from a completely different point of view. 

\section{Large $\Nf$ QCD}

Having outlined the critical point approach for $O(N)$ $\phi^4$ theory, we now 
concentrate on how one develops the same formalism for QCD in the large $\Nf$ 
limit. Examining the $d$-dimensional $\MSbar$ $\beta$-function for QCD, one 
finds that there is an equivalent Wilson-Fisher fixed point which can be 
accessed in powers of $1/\Nf$ as a function of $\epsilon$. However, to compute 
critical exponents in QCD in $d$-dimensions and relate them to the associated 
anomalous dimensions in perturbation theory at whatever order they are 
available, one does not compute with the QCD Lagrangian itself. Instead one 
exploits the properties of the $d$-dimensional fixed point by realising that at
this fixed point QCD is in the same universality class as the non-abelian 
Thirring model (NATM), \cite{31}. This is a four-fermi theory which is 
renormalizable in two dimensions and plays the role of the non-linear $\sigma$ 
model in the previous $\phi^4$ theory example. The Lagrangian of the NATM is 
\begin{equation}
L^{\mbox{\footnotesize{NATM}}} ~=~ i \bar{\psi}^{iI} \partialslash 
\psi^{iI} ~+~ \frac{\lambda^2}{2} \left( \bar{\psi}^{iI} T^a_{IJ} \gamma^\mu 
\psi^{iJ} \right)^2 
\end{equation}
where $\lambda$ is the coupling constant of the NATM and is dimensionless in
two dimensions. Introducing an {\em auxiliary} spin-$1$ field the Lagrangian
can be written as  
\begin{equation}
L^{\mbox{\footnotesize{NATM}}} ~=~ i \bar{\psi}^{iI} \partialslash 
\psi^{iI} ~+~ A_\mu^a \bar{\psi}^{iI} T^a_{IJ} \gamma^\mu \psi^{iJ} ~-~ 
\frac{A^{a \, 2}_\mu}{2\lambda} ~. 
\end{equation}
Clearly there is no field strength term in the non-abelian Thirring model and
from a critical point of view this is due to the fact that at this infrared
stable fixed point that operator is irrelevant. The critical behaviour is
driven by the common quark gluon vertex when one compares with the QCD
Lagrangian
\begin{equation} 
L^{\mbox{\footnotesize{QCD}}} ~=~ -~ \frac{1}{4} G_{\mu\nu}^a 
G^{a \, \mu\nu} ~-~ \frac{1}{2\tilde{\alpha}} (\partial^\mu A^a_\mu)^2 ~-~ 
\bar{c}^a \partial^\mu D_\mu c^a ~+~ i \bar{\psi}^{iI} \Dslash \psi^{iI}  
\end{equation}
where in this section we use $\tilde{\alpha}$ as the covariant gauge parameter
with $\tilde{\alpha}$~$=$~$0$ corresponding to the Landau gauge. Within the 
large $\Nf$ critical point formalism the triple and quartic gluon vertices are 
embedded inside quark loops with respectively three or four external auxiliary 
spin-$1$ fields, \cite{31}.

\begin{figure}[ht] 
\begin{center}
\scalebox{0.44}{\includegraphics{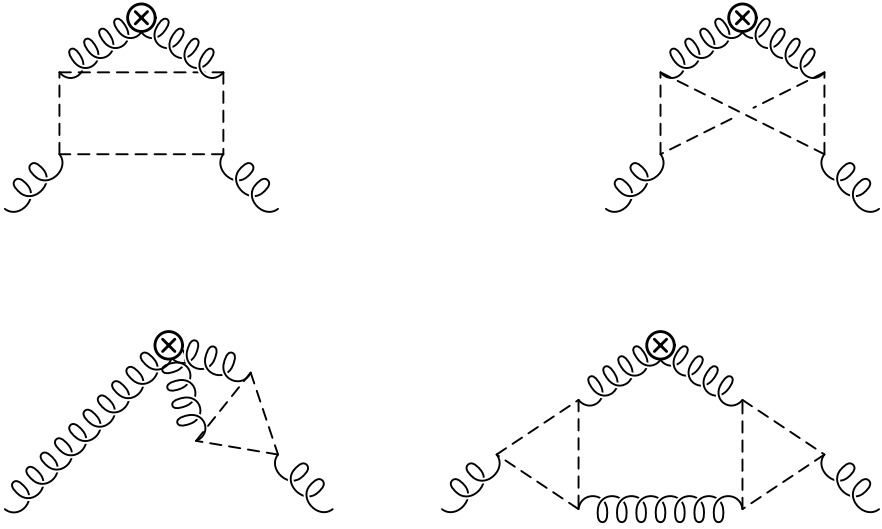}}
\caption{Feynman diagrams for $O(1/\Nf)$ correction to $\omega$.} 
\end{center}
\end{figure} 
As an example of the large $\Nf$ critical point formalism, the $O(1/\Nf)$
correction to the $\beta$-function can be computed by considering the critical
point structure of the gluon $2$-point function with the insertion of the
operator $G^a_{\mu\nu} G^{a \, \mu\nu}$, \cite{29}. The anomalous dimension of 
this operator will be related to the critical exponent of the associated 
coupling constant which is clearly the QCD $\beta$-function when expanded in 
powers of $\epsilon$ near four dimensions since 
$\omega$~$=$~$-$~$\beta^\prime(g_c)$. At criticality the propagators of the 
fields of the non-abelian Thirring model or QCD Lagrangian obey asymptotic 
scaling forms where the canonical exponents are derived from the dimensionality
of each term in the $d$-dimensional Lagrangian. Thus we have, in the Landau 
gauge, \cite{29,30}, 
\begin{equation}
\psi(k) ~\sim~ \frac{A\kslash}{(k^2)^{\mu-\alpha}} ~~,~~
A_{\mu\nu}(k) ~\sim~ \frac{B}{(k^2)^{\mu-\beta}}\left[ \eta_{\mu\nu}
- \frac{k_\mu k_\nu}{k^2} \right] 
\end{equation}
where $A$ and $B$ are momentum independent amplitudes and 
\begin{equation}
\alpha ~=~ \mu ~-~ 1 ~+~ \half \eta ~~~,~~~ 
\beta ~=~ 1 ~-~ \eta ~-~ \chi ~. 
\end{equation}
The exponents $\eta$ and $\chi$ are respectively the quark anomalous dimension
and the quark gluon vertex anomalous dimension at criticality. In large $\Nf$ 
there are four diagrams contributing to the gluon $2$-point function with the
operator inserted which are illustrated in Figure 1. The three loop diagram
is included since it is of the same order in large $\Nf$ as the two loop ones 
because in the large $\Nf$ counting one regards the amplitudes as 
$A$~$=$~$O(1)$ and $B$~$=$~$O(1/\Nf)$. These diagrams are infinite but can be 
regularized by shifting $\beta$ to $\beta$~$-$~$\Delta$ where $\Delta$ is the 
regularizing parameter, \cite{24,25,26}. It plays a role akin to $\epsilon$ in 
dimensional regularization and it is important to appreciate that within the
critical point large $\Nf$ method $\epsilon$ is not the regularization. The 
calculations are performed in {\em fixed} $d$-dimensions. After subtraction of 
the poles in $\Delta$, using the renormalization procedure of \cite{32}, the 
scaling behaviour of the remaining finite Green's function determines $\omega$ 
and we have, \cite{29},  
\begin{eqnarray}
\omega &=& (\mu - 2) ~-~ \left[ (2\mu-3)(\mu-3)C_F \right. \nonumber \\
&& \left. -~ \frac{(4\mu^4 - 18\mu^3 + 44\mu^2 - 45\mu + 14)C_A} 
{4(2\mu-1)(\mu-1)} \right] \frac{\eta_0}{T(R)N_{\! f}} 
\label{qcdom} 
\end{eqnarray}
where $\eta_0$ $=$ $(2\mu-1)(\mu-2)\Gamma(2\mu)/[4\Gamma^2(\mu) \Gamma(\mu+1) 
\Gamma(2-\mu)]$. The factors deriving from the group theory are defined by
\begin{equation}
T^a T^a ~=~ C_F I ~~~,~~~ f^{acd} f^{bcd} ~=~ C_A \delta^{ab} ~~~,~~~
\mbox{Tr} \left( T^a T^b \right) ~=~ T_F \delta^{ab} ~. 
\end{equation}
Consequently one can verify that the known leading $1/\Nf$ coefficients agree
with the $\MSbar$ four loop $\beta$-function of \cite{19}. Moreover, the 
unknown higher loop coefficients at $O(1/\Nf)$ are encoded in (\ref{qcdom}). 
For instance, if we write the leading large $\Nf$ part of the QCD 
$\beta$-function as
\begin{equation}
\beta(g) ~=~ \beta_0 g^2 ~+~ \sum_{n=1}^\infty a_{n+1} [T_F \Nf]^n g^{n+2}
\end{equation} 
then at {\em five} loops we have  
\begin{equation} 
a_5 ~=~ [(288\zeta(3) + 214)C_F + (480\zeta(3) - 229)C_A]/31104 ~. 
\end{equation} 

For $O(N)$ $\phi^4$ theory, it was possible with the critical point method to
go to several orders beyond the leading $1/N$ order. For QCD the same is
possible, in principle, but one has the added complication of the Lorentz
structure associated with fermions and the fact that the structure of the quark
gluon vertex is such that the method of uniqueness cannot be applied 
immediately. Instead one has to use integration by parts and other methods to
produce vertices which satisfy the uniqueness condition before that 
integration technique can be used. Nevertheless the quark anomalous dimension 
and quark mass anomalous dimensions have been determined at $O(1/\Nf^2)$. For 
the quark anomalous dimension in arbitrary covariant gauge we have, \cite{33}, 
\begin{equation}
\eta_1 ~=~ \frac{[(2\mu-1)(\mu-2) + \tilde{\alpha}\mu]C_F\eta_0}
{[(2\mu-1)(\mu-2)T_F]}
\end{equation}
and 
\begin{eqnarray} 
\eta_2 &=& \left[ \frac{}{} -~ \mu(2 \mu^2 + \mu \tilde{\alpha} - 5 \mu + 2) 
(\mu - 1) \left[ \bar{\Psi}^2(\mu) + \bar{\Phi}(\mu) \right] C_A \right. 
\nonumber \\ 
&& \left. ~~+~ (8 \mu^5 - 92 \mu^4 + 270 \mu^3 - 301 \mu^2 + 124 \mu 
- 12)( (2 \mu - 1)(\mu-2) + \mu \tilde{\alpha} ) \right. \nonumber \\
&& \left. ~~~~\quad\times ~ \frac{\bar{\Psi}(\mu) C_A }
{2 (2 \mu - 1) (2 \mu - 3) (\mu - 2)} \right. \nonumber \\ 
&& \left. ~~+~ 3\mu(\mu-1)[ \mu \tilde{\alpha} C_A + 2 (2 \mu^2 + \mu 
\tilde{\alpha} - 5 \mu + 2) C_F ] \hat{\Theta}(\mu) \right. \nonumber \\ 
&& \left. ~~-~ [(32 \mu^7 \tilde{\alpha} - 96 \mu^7 + 8 \mu^6 \tilde{\alpha}^2 
- 224 \mu^6 \tilde{\alpha} + 912 \mu^6 - 4 \mu^5 \tilde{\alpha}^3 - 84 \mu^5 
\tilde{\alpha}^2 \right. \nonumber \\
&& \left. ~~~~~~~ 
+~ 704 \mu^5 \tilde{\alpha} - 3360 \mu^5 
+ 16 \mu^4 \tilde{\alpha}^3 + 278 \mu^4 \tilde{\alpha}^2 
- 1124 \mu^4 \tilde{\alpha} \right. \nonumber \\
&& \left. ~~~~~~~ 
+~ 6240 \mu^4 - 19 \mu^3 \tilde{\alpha}^3 
- 387 \mu^3 \tilde{\alpha}^2 + 846 \mu^3 \tilde{\alpha} - 6292 \mu^3 \right. 
\nonumber \\
&& \left. ~~~~~~~+~ 6 \mu^2 \tilde{\alpha}^3 + 230 \mu^2 \tilde{\alpha}^2 
- 222 \mu^2 \tilde{\alpha} + 3416 \mu^2 - 48 \mu \tilde{\alpha}^2 \right. 
\nonumber \\
&& \left. ~~~~~~~-~ 4 \mu \tilde{\alpha} - 908 \mu + 88) C_A \mu \right. 
\nonumber \\
&& \left. ~~~~~~~-~ 8 (4 \mu^5 + 4 \mu^4 \tilde{\alpha} - 32 \mu^4 - 13 
\mu^3 \tilde{\alpha} + 75 \mu^3 + 8 \mu^2 \tilde{\alpha} - 70 \mu^2 
- 2 \mu \tilde{\alpha} \right. \nonumber \\
&& \left. ~~~~~~~~~~~~~+~ 32 \mu - 6) (2 \mu - 1) (2 \mu - 3) (\mu - 2) C_F ] 
\right. \nonumber \\
&& \left. ~~~~~~~/[4 (2 \mu - 1) (2 \mu - 3) (\mu - 1) (\mu - 2) \mu] \frac{}{}
\right] \frac{C_F \eta_0^2}{2(2\mu-1)^2(\mu-2)^2 T_F^2} 
\end{eqnarray} 
where
\begin{equation}
\eta ~=~ \sum_{i=1}^\infty \frac{\eta_i}{\Nf^i} ~.
\end{equation} 
The expression for the quark mass anomalous dimension is independent of 
$\tilde{\alpha}$ and in the same notation,  
\begin{equation} 
\eta_{\bar{\psi}\psi , 1} ~=~ -~ \frac{2C_F \eta_0}{(\mu-2)T_F} 
\end{equation}
and, \cite{33},  
\begin{eqnarray} 
\eta_{\bar{\psi}\psi , 2} &=& -~ \left[ 2 \left( 3\hat{\Theta}(\mu) 
+ \frac{(4\mu^3 - 13\mu^2 + 9\mu - 3)}{\mu^2(\mu-1)^2} \right) C_F \right. 
\nonumber \\  
&& ~\quad+~ \left. \left( \frac{(8\mu^5-92\mu^4+270\mu^3-301\mu^2+124\mu-12) 
\bar{\Psi}(\mu)}{2\mu(\mu-1)(2\mu-1)(2\mu-3)(\mu-2)} \right. \right. 
\nonumber \\ 
&& ~~~~~~~~~~~\quad\quad\quad-~ \left. \left. [16\mu^6 - 128\mu^5 + 480\mu^4  
- 900\mu^3 + 831\mu^2 \right. \right. \nonumber \\ 
&& \left. \left. ~~~~~~~~~~~~~~~~~~~~\quad\quad\quad\quad-~ 344\mu + 44)] 
/[4\mu(2\mu-1)(2\mu-3)(\mu-1)^2(\mu-2)] \right. \right. \nonumber \\ 
&& \left. \left. ~~~~~~~~~~~~~~~~\quad\quad\quad-~ \bar{\Psi}^2(\mu) ~-~ 
\bar{\Phi}(\mu) \frac{}{} \right) C_A \right] 
\frac{\mu(\mu-1)(2\mu-1)C_F\eta_0^2}{(2\mu-1)(\mu-2)^2 T_F^2} ~. 
\end{eqnarray} 

Having focused on the operator associated with the coupling constant of QCD, 
one can also repeat the same analysis for the analogous operator of the 
non-abelian Thirring model which is the dimension two gluon mass operator. 
Whilst it clearly is not a gauge invariant operator its anomalous dimension has
an interesting structure. Inserting $\half A^a_\mu A^{a\,\mu}$ into the same 
Green's function as $G^a_{\mu\nu} G^{a\,\mu\nu}$, the $O(1/\Nf)$ exponent in 
the Landau gauge is, \cite{34},  
\begin{equation}
\eta_{A^{2}}~=~-~\frac{C_{A}\eta _{1}^{\mbox{o}}}{4(\mu -2)T_{F}N_{\!f}} ~+~ 
O \left( \frac{1}{\Nf^2} \right) ~.  
\end{equation}
Interestingly this can be rewritten as 
\begin{equation}
\gamma_{A^2}(g_c) ~=~ \gamma_A(g_c) ~+~ \gamma_c(g_c)
\end{equation} 
in {\em all} dimensions $d$. It turns out that this is a general property of 
this operator in the Landau gauge. In \cite{35} it was shown that in QCD 
\begin{equation}
\gamma_{A^2}(g) ~=~ \gamma_A(g) ~+~ \gamma_c(g)
\end{equation} 
to all orders in perturbation theory having first been shown at three loops
in $\MSbar$ by explicit computation, \cite{36}. The explicit four loop value is
also now available in $SU(\Nc)$, \cite{37}. Although this operator is gauge 
variant, it has been the subject of intense study in recent years as an 
effective gluon mass term. See, for example, \cite{38}. Moreover, it has been 
studied in other gauges such as the maximal abelian gauge (MAG), 
\cite{39,40,18}. In this latter gauge the gauge field is written in terms of 
its diagonal (centre) and off-diagonal fields  
\begin{equation}
A^A_\mu T^A ~=~ A^a_\mu T^a ~+~ A^i_\mu T^i 
\label{asplit}
\end{equation} 
where $[ T^i, T^j]$ $=$ $0$, $T^i$ $\in$ $\{\mbox{centre}\}$,  
$1$~$\leq$~$i$~$\leq$~$\Nda$, $1$~$\leq$~$a$~$\leq$~$\Noda$ and 
$1$~$\leq$~$A$~$\leq$~$\NA$ with $\Nda$ and $\Noda$ the respective dimensions
of the centre of the group and its complement. (Here and in the next section we
use the index $A$ to denote the whole colour group, using the notation
usually used in discussing the maximal abelian gauge.) Then in this gauge it 
turns out that the off-diagonal BRST invariant mass operator  
\begin{equation}
O ~=~ \half A^a_\mu A^{a\,\mu} ~+~ \alpha \bar{c}^a c^a  
\label{magop} 
\end{equation} 
satisfies a Slavnov-Taylor identity, similar to that for the analogous operator
in the Landau gauge, which is, \cite{39},  
\begin{equation}
\gamma_{O}(g) ~=~ \gamma_{A^i}(g) ~- \gamma_{c^i}(g)
\end{equation} 
where $c^i$ is the ghost in the centre of the colour group. Whilst this
identity has been established on general grounds, for practical calculations,
such as studying the condensation of the operator $O$ of (\ref{magop}), one 
needs the {\em explicit} values of the anomalous dimensions. This requires the 
application of much of the earlier general discussion on renormalization and 
the maximal abelian gauge provides a comprehensive example to illustrate these 
general remarks.  

\section{QCD in non-linear gauges}

Given this interest in studying QCD in covariant gauges other than the 
canonical linear ones, we now focus on the explicit renormalization in the 
MAG. First, we define the gauge fixing for the maximal abelian gauge by 
recalling that for the Landau gauge. With (\ref{asplit}) the gauge fixing is 
given by 
\begin{equation} 
L^{\mbox{\footnotesize{Landau}}}_{\mbox{\footnotesize{gf}}} ~=~ \delta 
\bar{\delta} \left[ \frac{1}{2} A_\mu^A A^{A \, \mu} ~+~ \frac{1}{2} \alpha 
\bar{c}^A c^A \right] ~.  
\end{equation} 
By contrast, the maximal abelian gauge is such that one minimizes the quadratic
form in the off-diagonal sector only by fixing, \cite{39}, 
\begin{equation} 
L^{\mbox{\footnotesize{MAG}}}_{\mbox{\footnotesize{gf}}} ~=~ \delta 
\bar{\delta} \left[ \frac{1}{2} A_\mu^a A^{a \, \mu} ~+~ \frac{1}{2} \alpha 
\bar{c}^a c^a \right] ~+~ \delta \left[ \bar{c}^i \partial^\mu A_\mu^i \right] 
\label{maggf} 
\end{equation} 
where again $\delta$ and $\bar{\delta}$ are the respective BRST and anti-BRST
variations with  
\begin{eqnarray} 
\delta A^a_\mu &=& -~ \left( \partial_\mu c^a + g f^{ajc} A^j_\mu c^c
+ g f^{abc} A^b_\mu c^c + g f^{abk} A^b_\mu c^k \right) \nonumber \\ 
\delta c^a &=& g f^{abk} c^b c^k + \frac{1}{2} f^{abc} c^b c^c \nonumber \\
\delta \bar{c}^a &=& b^a ~~~,~~~ \delta b^a ~=~ 0 \nonumber \\
\delta A^i_\mu &=& -~ \left( \partial_\mu c^i + g f^{ibc} A^b_\mu c^c
\right) \nonumber \\
\delta c^i &=& \frac{1}{2} g f^{ibc} c^b c^c \nonumber \\
\delta \bar{c}^i &=& b^i ~~~,~~~ \delta b^i ~=~ 0 
\end{eqnarray} 
and 
\begin{eqnarray} 
\bar{\delta} A^a_\mu &=& -~ \left( \partial_\mu c^a + g f^{ajc} A^j_\mu c^c
+ g f^{abc} A^b_\mu c^c + g f^{abk} A^b_\mu c^k \right) \nonumber \\ 
\bar{\delta} c^a &=& -~ b^a + g f^{abc} c^b \bar{c}^c + g f^{abk} c^b \bar{c}^k
+ g f^{abk} \bar{c}^b c^k \nonumber \\
\bar{\delta} \bar{c}^a &=& g f^{abk} \bar{c}^b \bar{c}^k + \frac{1}{2} g f^{abc}
\bar{c}^b \bar{c}^c \nonumber \\
\bar{\delta} b^a &=& -~ g f^{abc} b^b \bar{c}^c - g f^{abk} b^b \bar{c}^k
+ g f^{abk} \bar{c}^b b^k \nonumber \\
\bar{\delta} A^i_\mu &=& -~ \left( \partial_\mu \bar{c}^i + g f^{ibc} 
A^b_\mu \bar{c}^c \right) ~~~,~~~ 
\bar{\delta} c^i ~=~ -~ b^i + g f^{ibc} c^b \bar{c}^c \nonumber \\
\bar{\delta} \bar{c}^i &=& \frac{1}{2} g f^{ibc} \bar{c}^b \bar{c}^c ~~~,~~~
\bar{\delta} b^i ~=~ -~ g f^{ibc} b^b \bar{c}^c ~.  
\end{eqnarray} 
The final term in (\ref{maggf}) is required to avoid a singular ghost 
propagator. The residual gauge freedom in the diagonal sector is fixed by
applying the usual Landau gauge. Though one can introduce an extra gauge
parameter, $\bar{\alpha}$, for the inversion to obtain a centre gluon
propagator. Varying $\delta$ and $\bar{\delta}$ gives contributions to the 
interaction Lagrangian with the remaining terms derived from
$(G^A_{\mu\nu})^2$. Thus, 
\begin{eqnarray}
L_{\mbox{\footnotesize{gf}}} &=& -~ \frac{1}{2\alpha} \left( \partial^\mu 
A^a_\mu \right)^2 - \frac{1}{2\bar{\alpha}} 
\left( \partial^\mu A^i_\mu \right)^2 + \bar{c}^a \partial^\mu \partial_\mu c^a
+ \bar{c}^i \partial^\mu \partial_\mu c^i \nonumber \\
&& +~ \left[ f^{abk} A^a_\mu \bar{c}^k \partial^\mu c^b 
- f^{abc} A^a_\mu \bar{c}^b \partial^\mu c^c 
- \frac{1}{\alpha} f^{abk} \partial^\mu A^a_\mu A^b_\nu A^{k \, \nu} 
\right. \nonumber \\
&& \left. ~~~~~~~  
-~ f^{abk} \partial^\mu A^a_\mu c^b \bar{c}^k 
- \frac{1}{2} f^{abc} \partial^\mu A^a_\mu \bar{c}^b c^c 
- 2 f^{abk} A^k_\mu \bar{c}^a \partial^\mu \bar{c}^b 
- f^{abk} \partial^\mu A^k_\mu \bar{c}^b c^c \right] g \nonumber \\  
&& +~ \left[ f_d^{acbd} A^a_\mu A^{b \, \mu} \bar{c}^c c^d  
- \frac{1}{2\alpha} f_o^{akbl} A^a_\mu A^{b \, \mu} A^k_\nu A^{l \, \nu} 
+ f_o^{adcj} A^a_\mu A^{j \, \mu} \bar{c}^c c^d \right. \nonumber \\
&& \left. ~~~~~~~~  
-~ \frac{1}{2} f_o^{ajcd} A^a_\mu A^{j \, \mu} \bar{c}^c c^d 
+ f_o^{ajcl} A^a_\mu A^{j \, \mu} \bar{c}^c c^l 
+ f_o^{alcj} A^a_\mu A^{j \, \mu} \bar{c}^c c^l \right. \nonumber \\
&& \left. ~~~~~~~~  
-~ f_o^{cjdi} A^i_\mu A^{j \, \mu} \bar{c}^c c^d  
- \frac{\alpha}{4} f_d^{abcd} \bar{c}^a \bar{c}^b c^c c^d  
- \frac{\alpha}{8} f_o^{abcd} \bar{c}^a \bar{c}^b c^c c^d  
\right. \nonumber \\
&& \left. ~~~~~~~~  
+~ \frac{\alpha}{8} f_o^{acbd} \bar{c}^a \bar{c}^b c^c c^d 
- \frac{\alpha}{4} f_o^{abcl} \bar{c}^a \bar{c}^b c^c c^l  
+ \frac{\alpha}{4} f_o^{acbl} \bar{c}^a \bar{c}^b c^c c^l  
\right. \nonumber \\
&& \left. ~~~~~~~~  
-~ \frac{\alpha}{4} f_o^{albc} \bar{c}^a \bar{c}^b c^c c^l  
+ \frac{\alpha}{2} f_o^{akbl} \bar{c}^a \bar{c}^b c^k c^l \right] g^2 
\label{maglag} 
\end{eqnarray}   
where $f_d^{ABCD}$ $=$ $f^{ABi} f^{CDi}$ and $f_o^{ABCD}$ $=$ 
$f^{ABe} f^{CDe}$. Although this is a much more involved Lagrangian than that
which results in the Landau gauge where there are a handful of interactions, it
can be shown that it contains the usual covariant gauge fixed Lagrangian,
\cite{39}. Moreover, in the limit where the fields deriving from the centre of
the colour group are set to zero one recovers the non-linear Curci-Ferrari
gauge, \cite{41}. In other words the off-diagonal sector of the maximal abelian
gauge corresponds to QCD fixed in the Curci-Ferrari gauge. Clearly the 
Lagrangian includes quartic ghost interactions which are always a feature of a 
non-linear gauge fixing. Though they do not destroy the renormalizability of 
the maximal abelian gauge Lagrangian which has been established by the 
algebraic renormalization technology, \cite{39,40}. This is crucial to 
formulating the renormalization of QCD in the maximal abelian gauge as it 
produces the Slavnov-Taylor identities originating from the BRST 
transformations. More importantly it determines the {\em form} of the 
renormalization of the fields and parameters as  
\begin{eqnarray} 
A^{a \, \mu}_{\mbox{\footnotesize{o}}} &=& \sqrt{Z_A} \, A^{a \, \mu} ~~,~~ 
A^{i \, \mu}_{\mbox{\footnotesize{o}}} ~=~ \sqrt{Z_{A^i}} \, A^{i \, \mu} 
\nonumber \\ 
c^a_{\mbox{\footnotesize{o}}} &=& \sqrt{Z_c} \, c^a ~~~,~~~ 
\bar{c}^a_{\mbox{\footnotesize{o}}} ~=~ \sqrt{Z_c} \, \bar{c}^a \nonumber \\
c^i_{\mbox{\footnotesize{o}}} &=& \sqrt{Z_{c^i}} \, c^i ~~~,~~ 
\bar{c}^i_{\mbox{\footnotesize{o}}} ~=~ \frac{\bar{c}^i}{\sqrt{Z_{c^i}}}
\nonumber \\
\psi_{\mbox{\footnotesize{o}}} &=& \sqrt{Z_\psi} \psi ~~~,~~~  
g_{\mbox{\footnotesize{o}}} ~=~ \mu^\epsilon Z_g \, g \nonumber \\ 
\alpha_{\mbox{\footnotesize{o}}} &=& Z^{-1}_\alpha Z_A \, \alpha ~~~,~~~  
\bar{\alpha}_{\mbox{\footnotesize{o}}} ~=~ Z^{-1}_{\alpha^i} Z_{A^i} \, 
\bar{\alpha} ~.  
\end{eqnarray} 
Consequently, one observes that the diagonal ghost $2$-point function is
finite. If this had not been deduced via algebraic renormalization, then in an 
explicit renormalization of the diagonal ghost $2$-point function one would 
have erroneoulsy deduced that $Z_{c^i}$ was unity. Instead to determine the 
correct $Z_{c^i}$ one has to renormalize a Green's function containing one 
$Z_{c^i}$ where all the other associated renormalization constants are known. 
For this case one choice would be the $A^a_\mu c^b \bar{c}^i$ vertex
renormalization. Another useful feature is that the diagonal gluon wave
function renormalization constant is not independent but is in fact related to 
that of the coupling constant. Hence, the $\beta$-function can be determined 
without resort to the renormalization of a vertex function. Moreover, this
simplification is similar to what happens in QCD fixed in the background field 
gauge, \cite{42}. Though in practical calculations it reduces substantially the 
number of diagrams to be determined as well as computation time which can be 
significant at high loop orders. The key point is that the preliminary 
analysis from algebraic renormalization not only determines the structure of
the renormalization constants consistent with the underlying symmetry, but also
provides an efficient route for computing the anomalous dimensions themselves. 

Given that the maximal abelian gauge gluon mass operator does not renormalize 
independently, to have its explicit anomalous dimension requires only knowledge
of the diagonal ghost anomalous dimension. Clearly given the large number of 
interactions in (\ref{maglag}), a three loop renormalization can only proceed 
with an automatic computer algebra approach for which the {\sc Mincer} 
algorithm is the ideal tool, using the {\sc Form} version, \cite{16}, and 
{\sc Qgraf}, \cite{20}. In completing the full three loop maximal abelian gauge
renormalization of QCD, $37322$ Feynman diagrams had to be considered, 
\cite{18}. Though not all of these are non-zero. Some vanish trivially by the 
group theory structure of a diagram. For instance, where one has a self-energy 
insertion in a propagator line with a diagonal and off-diagonal field as 
external to this subgraph, then it will vanish due to
\begin{equation}
f^{ijk} ~=~ 0 ~~~,~~~ f^{ijc} ~=~ 0 ~~~,~~~ 
f^{ibc} ~\neq~ 0 ~~~,~~~ f^{abc} ~\neq~ 0 ~. 
\end{equation}  
The result of the renormalization procedure yields the anomalous dimensions,
\cite{18}, which have the structure illustrated by the example, 
\begin{eqnarray} 
\gamma_{O}(a) &=& \frac{1}{12 \Noda} \left[ \Noda \left( ( -~ 3 \alpha 
+ 35 ) C_A - 16 T_f \Nf \right) + \Nda \left( ( -~ 6 \alpha - 18 ) C_A \right) 
\right] a \nonumber \\
&& +~ \frac{1}{96 {\Noda}^2} \left[ {\Noda}^2 \left(  ( -~ 6 \alpha^2 
- 66 \alpha + 898 ) C_A^2 - 560 C_A T_f \Nf 
\right. \right.  \nonumber \\
&& \left. \left. ~-~ 384 C_F T_f \Nf \right) 
+~ \Noda \Nda \left(  ( -~ 54 \alpha^2 - 354 \alpha 
- 323 ) C_A^2 + 160 C_A T_f \Nf \right) \right. \nonumber \\
&& \left. ~+~ {\Nda}^2 \left(  ( -~ 60 \alpha^2 - 372 \alpha 
+ 510 ) C_A^2 \right) \right] a^2 \nonumber \\
&& +~ \frac{1}{6912 {\Noda}^3} \left[ {\Noda}^3 ( ( -~ 162 \alpha^3 
- 2727 \alpha^2 - 2592 \zeta(3) \alpha - 18036 \alpha
\right. \nonumber \\
&& \left. ~
-~ 1944 \zeta(3) 
+ 302428 ) C_A^3 + ( 6912 \alpha + 62208 \zeta(3) - 356032 ) C_A^2 T_F \Nf 
\right. \nonumber \\
&& \left. ~
+~ ( -~ 82944 \zeta(3) 
- 79680 ) C_A C_F T_F \Nf + 49408 C_A T_F^2 \Nf^2 
\right. \nonumber \\
&& \left. ~
+~ 13824 C_F^2 T_F \Nf 
+ 33792 C_F T_F^2 \Nf^2 )
\right. \nonumber \\
&& \left. ~
+~ {\Noda}^2 {\Nda} ( ( -~ 2754 \alpha^3 + 648 \alpha^2 \zeta(3)
- 28917 \alpha^2 
- 4212 \alpha \zeta(3) 
- 69309 \alpha
\right. \nonumber \\
&& \left. ~
+~ 37260 \zeta(3) - 64544 ) C_A^3
+ ( 25488 \alpha + 103680 \zeta(3) 
\right. \nonumber \\
&& \left. ~
-~ 13072 ) C_A^2 T_F \Nf 
+ ( -~ 165888 \zeta(3) + 155520 ) C_A C_F T_F \Nf 
\right. \nonumber \\
&& \left. ~
+~ 17920 C_A T_F^2 \Nf^2 ) 
+ {\Noda} {\Nda}^2 (  ( -~ 7884 \alpha^3 + 22680 \alpha^2 \zeta(3)
- 84564 \alpha^2 
\right. \nonumber \\
&& \left. ~
+~ 97524 \alpha \zeta(3) - 47142 \alpha 
+ 433836 \zeta(3) - 56430 ) C_A^3
\right. \nonumber \\
&& \left. ~
+~ ( 25056 \alpha - 124416 \zeta(3) 
- 18144 ) C_A^2 T_F \Nf ) 
\right. \nonumber \\
&& \left. ~
+~ {\Nda}^3 (  ( -~ 6480 \alpha^3 + 34992 \alpha^2 \zeta(3) 
- 70092 \alpha^2 + 8424 \alpha \zeta(3) 
\right. \nonumber \\
&& \left. ~
+~ 114912 \alpha 
+ 77112 \zeta(3) - 161028 ) C_A^3) \right] a^3 ~+~ O(a^4) 
\end{eqnarray} 
where $a$~$=$~$g^2/(16\pi^2)$ is conventionally used as the coupling constant.
To be confident that the anomalous dimensions are in fact correct, the standard
internal renormalization group checks we discussed above have been shown to 
hold. Moreover, since the $\beta$-function is a gauge independent object then 
the cancellation of $\alpha$ and the parameters $\Nda$ and $\Noda$, provided 
not only a strong check on the calculation but on the group theory {\sc Form}
module which handles the consequences of splitting the gauge group into two
sectors. Additional checks for this specific computation are provided by noting
that the anomalous dimensions of the off-diagonal fields are correctly 
equivalent for all $\alpha$ to those of QCD fixed in the non-linear 
Curci-Ferrari gauge, which was introduced in \cite{41}, in the limit 
$\Nda/\Noda$~$\rightarrow$~$0$ where the explicit results were given in 
\cite{43,36}.

\section{Renormalization and non-locality}

We now turn to more recent aspects of renormalization and that is the
renormalization of quantum field theories where a non-locality is present. The 
general properties of renormalization theory as discussed, for example, in 
\cite{44}, is based on the assumption that the Lagrangian is local. In other 
words there are no non-local interactions or operators. However, in certain 
field theories of interest, such as QCD, one encounters important operators 
which are non-local but whose properties require investigation. We now 
summarize the status of two such studies where although there is a degree of 
non-locality present, it falls into a class which does not prevent calculations
from being performed. By this we mean that the non-local operators can be 
rewritten in terms of a {\em finite} number of local fields and operators. We 
refer to this as a localizable non-locality. By contrast, there are operators 
which, whilst being non-local, do not allow for a {\em finite} number of 
auxiliary fields to lead to a local Lagrangian. We term this a non-localizable 
non-locality. 

The first example is that of the Gribov problem which relates to the
impossibility of {\em globally} fixing the gauge in a non-abelian gauge theory. 
This was first pointed out by Gribov, \cite{45}, who proceeded to construct a 
path integral to study the problem. There the non-abelian gauge field is
restricted to the region defined by the first zero of the Faddeev-Popov
operator. This region, known as the Gribov volume, is of finite size and
characterised by the Gribov mass, $\gamma$. The resulting path integral,
\cite{45},  
\begin{equation}
Z ~=~ \int { D} A ~ \delta ( \partial^\mu A^a_\mu) \det 
\left( - \partial^\nu D^a_\nu \right) ~ e^{- S} 
\end{equation}
essentially leads to a non-local Lagrangian 
\begin{equation} 
L ~=~ -~ \frac{1}{4} G_{\mu\nu}^a G^{a \, \mu\nu} ~+~ \frac{C_A\gamma^4}{2} 
A^a_\mu \, \frac{1}{\partial^\nu D_\nu} A^{a \, \mu} ~-~ 
\frac{d \NA \gamma^4}{2g^2} ~.  
\end{equation} 
The parameter $\gamma$ is not independent and satisfies the Gribov gap 
equation which is, at one loop, \cite{45}, 
\begin{equation} 
1 ~=~ C_A \left[ \frac{5}{8} ~-~ \frac{3}{8} \ln \left( 
\frac{C_A\gamma^4}{\mu^4} \right) \right] a ~+~ O(a^2) ~. 
\end{equation}  
Subsequently, in \cite{46,47} Zwanziger managed to localize the original 
Gribov path integral by introducing a set of extra ghost fields 
$\{\phi^{ab}_\mu, \bar{\phi}^{ab}_\mu, \omega^{ab}_\mu, \bar{\omega}^{ab}_\mu 
\}$ where the last two fields are anti-commuting. This led to the 
Gribov-Zwanziger Lagrangian in the Landau gauge, \cite{46,47},  
\begin{eqnarray}
L^{\mbox{\footnotesize{GZ}}} &=& L^{\mbox{\footnotesize{QCD}}} ~+~ 
\bar{\phi}^{ab \, \mu} \partial^\nu
\left( D_\nu \phi_\mu \right)^{ab} ~-~ \bar{\omega}^{ab \, \mu} \partial^\nu 
\left( D_\nu \omega_\mu \right)^{ab} \nonumber \\  
&& -~ g f^{abc} \partial^\nu \bar{\omega}^{ae}_\mu \left( D_\nu c \right)^b
\phi^{ec \, \mu} \nonumber \\
&& -~ \frac{\gamma^2}{\sqrt{2}} \left( f^{abc} A^{a \, \mu} \phi^{bc}_\mu ~+~ 
f^{abc} A^{a \, \mu} \bar{\phi}^{bc}_\mu \right) ~+~ 
\frac{d \NA \gamma^4}{2g^2} 
\label{laggz} 
\end{eqnarray} 
which clearly involves only a finite number of local interactions. Not only is
the Lagrangian local, it is also renormalizable, \cite{48,49}, which implies 
that one can perform loop computations and study the implications the Gribov 
parameter has on the infrared structure of QCD. The renormalization structure 
of (\ref{laggz}) is interesting in that in the Landau gauge the renormalization
constants of the extra ghost fields are not independent with, \cite{48,49}, 
\begin{equation}
Z_\phi ~=~ Z_\omega ~=~ Z_c ~.
\end{equation} 
Moreover, the renormalization of the Gribov parameter is not independent in the
Landau gauge, satisfying, \cite{49}, 
\begin{equation}
\gamma_\gamma(a) ~=~ \frac{1}{4} \left[ \gamma_A(a) ~+~ \gamma_c(a) \right] 
\end{equation} 
which is similar to the renormalization of $\half A^a_\mu A^{a \, \mu}$.
Moreover, the quark, gluon, Faddeev-Popov ghost anomalous dimensions and the
$\beta$-function are unaltered by the presence of $\gamma$ and the extra
Zwanziger ghost fields in $\MSbar$. In a general linear covariant gauge one has
the additional two loop $\MSbar$ results for the anomalous dimensions that   
\begin{eqnarray} 
\gamma_\phi(a) &=& \gamma_\omega(a) ~=~ \gamma_c(a) \nonumber \\  
\gamma_\gamma(a) &=& ( 16 T_F \Nf - ( 35 + 3 \alpha )C_A ) \frac{a}{48} 
\nonumber \\
&& +~ ( 192 C_F T_F \Nf + 280 C_A T_F \Nf - ( 449 - 3 \alpha ) ) 
\frac{a^2}{192} ~+~ O(a^3) 
\end{eqnarray} 
which was deduced using {\sc Mincer}. 

Equipped with these properties one can compute the corrections to the one loop
mass gap equation which Gribov originally derived. This corresponds to the 
horizon condition definition of \cite{47}, which is equivalent to 
\begin{equation} 
f^{abc} \langle A^{a \, \mu}(x) \phi^{bc}_\mu(x) \rangle ~=~ 
\frac{d \NA \gamma^2}{\sqrt{2}g^2} 
\end{equation}
which can be evaluated order by order in perturbation theory using the
propagators, \cite{46,47,50}, 
\begin{eqnarray}
\langle A^a_\mu(p) A^b_\nu(-p) \rangle &=& -~ 
\frac{\delta^{ab}p^2}{[(p^2)^2+C_A\gamma^4]} P_{\mu\nu}(p) \nonumber \\  
\langle A^a_\mu(p) \bar{\phi}^{bc}_\nu(-p) \rangle &=& -~ 
\frac{f^{abc}\gamma^2}{\sqrt{2}[(p^2)^2+C_A\gamma^4]} P_{\mu\nu}(p) 
\nonumber \\  
\langle \phi^{ab}_\mu(p) \bar{\phi}^{cd}_\nu(-p) \rangle &=& -~ 
\frac{\delta^{ac}\delta^{bd}}{p^2}\eta_{\mu\nu} ~+~  
\frac{f^{abe}f^{cde}\gamma^4}{p^2[(p^2)^2+C_A\gamma^4]} P_{\mu\nu}(p) 
\end{eqnarray} 
where the gluon propagator is suppressed in the infrared. Hence, at two loops
in $\MSbar$, \cite{50},  
\begin{eqnarray} 
1 &=& C_A \left[ \frac{5}{8} - \frac{3}{8} \ln \left( 
\frac{C_A\gamma^4}{\mu^4} \right) \right] a \nonumber \\ 
&& +~ \left[ C_A^2 \left( \frac{2017}{768} - \frac{11097}{2048} s_2
+ \frac{95}{256} \zeta(2)
- \frac{65}{48} \ln \left( \frac{C_A\gamma^4}{\mu^4} \right) \right. \right. 
\nonumber \\ 
&& \left. \left. ~~~~~~~~~~~~+~ \frac{35}{128} \left( \ln \left( 
\frac{C_A\gamma^4}{\mu^4} \right) \right)^2 + \frac{1137}{2560} \sqrt{5} 
\zeta(2) - \frac{205\pi^2}{512} \right) \right. \nonumber \\
&& \left. ~~~~~~~~~+~ C_A T_F \Nf \left( -~ \frac{25}{24} - \zeta(2)
+ \frac{7}{12} \ln \left( \frac{C_A\gamma^4}{\mu^4} \right) \right. \right.
\nonumber \\ 
&& \left. \left. ~~~~~~~~~~-~ \frac{1}{8} \left( \ln \left( 
\frac{C_A\gamma^4}{\mu^4} \right) \right)^2 + \frac{\pi^2}{8} \right) \right] 
a^2 +~ O(a^3) 
\end{eqnarray} 
where $s_2$ $=$ $(2\sqrt{3}/9) \mbox{Cl}_2(2\pi/3)$. The one and two loop
corrections are evaluated using the vacuum bubble approach discussed earlier.
Though we note that the basic two loop multiscale vacuum bubble integral
$I(m_1^2,m_2^2,m_3^2)$, where 
\begin{equation}
I(m_1^2,m_2^2,m_3^2) ~=~ 
\int_{kl} \frac{1}{[k^2-m_1^2][l^2-m_2^2][(k-l)^2-m_3^2]}  
\end{equation}
has to be determined for $m_i^2$~$\in$~$\{0, i\sqrt{C_A}\gamma^2, 
-i\sqrt{C_A}\gamma^2\}$. For instance, the quantities $s_2$ and 
$\sqrt{5}\zeta(2)$ arise from the finite parts of  $I(i\sqrt{C_A}\gamma^2, 
i\sqrt{C_A}\gamma^2, i\sqrt{C_A}\gamma^2)$ and $I(i\sqrt{C_A}\gamma^2, 
i\sqrt{C_A}\gamma^2, - i\sqrt{C_A}\gamma^2)$ respectively. 

One interesting consequence of this gap equation is that it ensures the 
Kugo-Ojima confinement condition, \cite{51}, of Faddeev-Popov ghost propagator 
enhancement is preserved at two loops. If one considers the full ghost 
propagator to have the form  
\begin{equation}
G_c(p^2) ~=~ \frac{\delta^{ab}}{p^2[ 1 + u(p^2) ]}
\end{equation}
where $u(p^2)$ represents the radiative corrections, then the Kugo-Ojima
confinement criterion is that $u(0)$~$=$~$-$~$1$, \cite{51}. Hence, the ghost 
propagator behaves as $1/(p^2)^2$ as $p^2$~$\rightarrow$~$0$. As $u(p^2)$ 
corresponds to the loop corrections of the ghost $2$-point function, then 
computing $u(p^2)$ in the vacuum bubble expansion to two loops one can check if
the Kugo-Ojima criterion holds at this order. Using the Gribov-Zwanziger
Lagrangian, (\ref{laggz}), it transpires that the gap equation emerges as the
$u(0)$~$+$~$1$ term at $O(p^2)$ so that $u(0)$~$=$~$-$~$1$ precisely at two
loops, \cite{50}.

Finally, we discuss a more recent study of non-locality in QCD, \cite{52}. 
Earlier we considered the gauge variant dimension two operator 
$\half A^a_\mu A^{a\,\mu}$ given the current interest in it as a potential
mass operator for the gluon. However, if one wishes to have a gauge invariant
Lagrangian with gluon mass then this operator is excluded. Instead to preserve
gauge invariance and insist on a mass operator, one has to allow for a
non-local mass operator. In \cite{52}, it was pointed out that aside from the
one usually associated with the (non-renormalizable) Stueckelberg term,
\begin{equation}
\tilde{A}^2_\mu ~=~  
\stackrel{\mbox{min}}{\mbox{\begin{tiny}$\{U\}$\end{tiny}}} 
\int \left( A^U_\mu \right)^2 
\end{equation}
there is another independent operator which is  
\begin{equation}
G^a_{\mu\nu} \frac{1}{D^2} G^{a\,\mu\nu} 
\end{equation}
which can be added to the gauge fixed Lagrangian as 
\begin{equation}
L ~=~ L^{\mbox{\footnotesize{gf}}} ~-~ \frac{m^2}{4} G^a_{\mu\nu} 
\frac{1}{D^2} G^{a\,\mu\nu} ~. 
\end{equation}
This is a localizable operator in the sense we defined earlier. Consequently,
\cite{52}, it can be localized by introducing the ghost fields $\{B^a_{\mu\nu},
\bar{B}^a_{\mu\nu},H^a_{\mu\nu},\bar{H}^a_{\mu\nu}\}$ where the last two are
anti-commuting, to give  
\begin{equation}
L ~=~ L^{\mbox{\footnotesize{gf}}} ~+~ \frac{im}{4} ( B^a_{\mu\nu} ~-~ 
\bar{B}^a_{\mu\nu} ) G^{a\,\mu\nu} ~+~ \frac{1}{4} \bar{B}^a_{\mu\nu}
\left( D^\sigma D_\sigma B^{\mu\nu} \right)^a ~-~ \frac{1}{4} 
\bar{H}^a_{\mu\nu} \left( D^\sigma D_\sigma H^{\mu\nu} \right)^a ~. 
\label{lagnl}
\end{equation}
This Lagrangian has been analysed by algebraic renormalization and whilst it
is renormalizable, it is not multiplicatively renormalizable since new quartic
ghost terms are generated through quantum corrections, \cite{52}. Moreover, the 
localized dimension three operator $( B^a_{\mu\nu}$~$-$~$\bar{B}^a_{\mu\nu} ) 
G^{a\,\mu\nu}$ operator itself mixed into the lower dimensional mass operators
$\left( \bar{B}^a_{\mu\nu} B^{a\,\mu\nu} \right.$~$-$~$\left. 
\bar{H}^a_{\mu\nu} H^{a\,\mu\nu} \right)$ and 
$(\bar{B}^a_{\mu\nu}-B^a_{\mu\nu})^2$. Given that (\ref{lagnl}) is 
renormalizable, the one loop anomalous dimensions have been determined. Clearly
the addition of such an operator ought not to affect the established anomalous 
dimensions of the gluon, Faddeev-Popov ghost and quark fields and explicit 
calculations verify this. Essentially within the respective $2$-point functions
the contributions from the extra ghost fields cancel. Moreover,
\begin{equation}
\gamma_B(a) ~=~ \gamma_H(a) ~=~ ( \alpha - 3 ) C_A a ~+~ O(a^2) 
\end{equation}
and the anomalous dimension of the mass operator itself can be deduced by
inserting the gauge invariant equivalent operator 
$(B^a_{\mu\nu}$~$-$~$\bar{B}^a_{\mu\nu}) G^{a\,\mu\nu}$ into the 
$B^a_{\mu\nu}$-gluon $2$-point function. The explicit {\sc Mincer} calculation
yields the result, \cite{52},  
\begin{equation}
\gamma_{B G}(a) ~=~ -~ \left( \frac{11}{6} C_A - \frac{2}{3} T_F \Nf \right)
a^2 ~+~ O(a^3) 
\label{nlad} 
\end{equation}
which is not only independent of the gauge parameter as it ought to be, but is
equivalent to the one loop QCD $\beta$-function. This is the same as the one
loop anomalous dimension of higher dimensional operators with similar Lorentz
structure, \cite{53}. In other words the operators $G^a_{\mu\nu} 
G^{a\,\mu\nu}$, $D_\mu G^a_{\nu\sigma} D^\mu G^{a\,\nu\sigma}$,  
$D_\mu D_\nu G^a_{\sigma\rho} D^\mu D^\nu G^{a\,\sigma\rho}$ and  
$D_\mu D_\nu D_\sigma G^a_{\rho\theta} D^\mu D^\nu D^\sigma G^{a\,\rho\theta}$ 
all have {\em same} one loop anomalous dimension as the non-local operator. We
complete this section by remarking that it would be interesting to determine
the higher loop corrections to (\ref{nlad}). 

\section{Conclusions} 

We close with general remarks. First, in the initial part of the article we
have reviewed the techniques and general structure associated with the 
renormalization of quantum field theories in the context of $O(N)$ $\phi^4$
theory and QCD. One of the interesting aspects of renormalization is the rich
interplay between the abstract and mundane exercise of evaluating 
renormalization group functions and the implications such results have on the
underlying physics. For example, the anomalous dimensions when evaluated at
a critical point of the $\beta$-function lead to predictions for the scaling
behaviour of Green's functions which can be measured in experiments. However,
from a technical point of view the critical exponents, when studied in the
large $N$ expansion, also actually complement the checking of the explicit
perturbative results. For large loop order computations this plays a role in 
establishing the correctness of the result as well as providing {\em new}
information on the as yet undetermined terms in the series at several orders
in large $N$. This can be important due to the fact that such high order
computations can presently only be undertaken with intense use of computer
algebra and symbolic manipulation techniques. The latter part of the article 
dealt with the application of renormalization to more current computations 
including the renormalization of QCD in non-linear gauges. Although this 
follows the application of established techniques, such calculations do provide
additional checks on the already determined three loop gauge independent 
renormalization group functions such as the $\beta$-function. Whilst much of 
the theory of renormalization is based on the assumption of locality of the 
initial Lagrangian, we have also touched on very recent calculations of 
operators which are {\em non-local} in structure. Although these fall in the 
class of localizable non-local operators, it is possible that such studies 
might open the possibility of studying problems in QCD where such operators are
important in probing the infrared structure of the theory. This is more 
evident, for example, in the Gribov problem where the structure of the ghost 
propagators satisfies the Kugo-Ojima confinement condition at {\em two} loops.  

\section*{Acknowledgement} The author thanks the Fields Institute for 
financial support in attending the meeting.

\end{document}